\documentclass[9pt,twocolumn,twoside]{opticajnl}
\journal{opticajournal} 

\setboolean{shortarticle}{true}

\usepackage{lineno}
\usepackage[normalem]{ulem}

\title{The role of the inverse Cherenkov effect in the formation of ultrashort Raman solitons in silica microspheres}

\author[1, *]{Alexey N. Osipov}
\author[2, 3]{Elena A. Anashkina }
\author[1]{Alexey V. Yulin}

\affil[1]{Department of Physics, ITMO University, Saint Petersburg 197101, Russia}
\affil[2]{A.V. Gaponov-Grekhov Institute of Applied Physics of the Russian Academy of Sciences, 46 Ulyanov Street, 603950 Nizhny Novgorod, Russia}
\affil[3]{Advanced School of General and Applied Physics, Lobachevsky State University of Nizhny Novgorod, 23 Gagarin Ave., 603022 Nizhny Novgorod, Russia}

\affil[*]{aleksey.osipov@metalab.ifmo.ru}

\begin{abstract}
We theoretically demonstrate a new regime of the formation of ultrashort optical solitons in spherical silica microresonators with whispering gallery modes.The solitons are driven by a coherent CW pump at the frequency in the range of normal dispersion, and the energy is transferred from this pump to the solitons via two channels: Raman amplification and the inverse Cherenkov effect. We discuss three different regimes of soliton propagation and we also show that these Raman solitons can be controlled by weak coherent CW signals. 
\end{abstract}

\setboolean{displaycopyright}{false} 

\begin{document}

\maketitle

Chip-scale sources of optical frequency combs (microcombs) have been of great interests to the scientific community over the last decade \cite{ herr2014temporal, obrzud2017temporal, kippenberg2018dissipative, yi2015soliton, joshi2016thermally, brasch2016photonic, Dolinina}. 
The physical origin of these microcombs is often the formation of dissipative Kerr solitons (DKS).
The interest in the dynamics of DKS and other types of microcombs is not only fundamental, but also motivated by their  applications 
in spectroscopy \cite{kippenberg2018dissipative}, telecommunications \cite{marin2017microresonator}, artificial intelligence \cite{xu202111}, astronomy \cite{obrzud2019microphotonic}, and distance measurements \cite{suh2018soliton}.

In dissipative systems, solitons can only survive in the presence of an external pump that compensates for  energy losses. The energy balance can be provided by a coherent resonant pump directly exciting cavity modes. Another way to   supply energy to the soliton is to use a pump, which produces an effective gain for some resonator modes. The latter can be achieved by various means, including Raman amplification. The solitons supported by Raman gain, the Raman solitons or Stokes solitons, have recently been the subject of intense research and are found in various optical systems including fiber cavities \cite{li2023ultrashort},
and silica microresonators of different shapes \cite{yang2017stokes, anashkina2022thermo, Do2023OpEx}. 

The mechanism of the Raman soliton formation can roughly be seen as the compensation of anomalous dispersion by Kerr nonlinearity, as it occurs in Schrödinger solitons, and the compensation of the losses by the Raman amplification. Therefore, the properties of Stokes (Raman) solitons depend dramatically on the resonator dispersion. Moreover, the presence of higher-order dispersion not only affects the soliton shape, but can also give rise to the resonance radiation known as Cherenkov one \cite{PhysRevE.72.016619, cherenkov2017dissipative}. This radiation enriches the radiation spectrum and thus plays an important role in the frequency comb generation.

Raman solitons can be excited by a CW holding beam with the frequency lying in the normal dispersion range and producing the Raman gain in the range of anomalous dispersion. In the presence of focusing Kerr nonlinearity, the modulation instability may occur, leading to splitting the waves produced by the Raman gain into trains of solitons \cite{anashkina2022thermo}.  The central frequency of these Raman solitons is, typically, close to the frequency where the Raman gain is maximal. 

The dynamics of the Raman soliton and the importance of the phase and group velocities matching conditions in a pulsed-pumped fiber cavity are discussed in \cite{li2023ultrashort}, where it is shown that the phase locking greatly increases the stability of the generated frequency comb. To explain the mechanism of the group velocity locking the system with CW pump is considered in \cite{li2024continuous}.

The purpose of the present paper is to reveal the importance of the Cherenkov resonance for the formation of Raman solitons. Indeed, in the presence of higher-order dispersion the Raman solitons, very much like conservative solitons in fibers,  can emit dispersive waves via Cherenkov mechanism. This radiation can easily be detected by a narrow peak in the spectrum. 
For some parameters the frequency of the Cherenkov resonance can coincide with the frequency of the pump. Then the Cherenkov radiation will either constructively or destructively interfere with the field directly generated by the coherent pump. This opens a new channel of the energy exchange between the soliton and the pump via the inverse Cherenkov effect \cite{PhysRevA.98.023833}.

This synchronism not only opens a new channel of the energy transfer from the pump to the soliton, but also locks the phase of the soliton to the phase of the pump. In this paper, we report the existence of the synchronized (phase-locked to the pump) and non-synchronized (whose phase is not locked to the pump phase)  Raman solitons in silica microspheres with whispering gallery modes pumped by a CW holding beam.
The advantage of these systems is their compactness, ease of fabrication, dispersion tailoring (by controlling diameter during fabrication), and a very high quality factor (10$^7$-10$^8$) which allows to use the pumps of relatively low power (of the order of 10 mW). Note that non-synchronised Raman solitons in microspheres have already been observed experimentally \cite{anashkina2022thermo},  while synchronized solitons predicted in this Letter have yet to be attained. 

\begin{figure}[!t]
\centering
\includegraphics[width=0.9\linewidth]{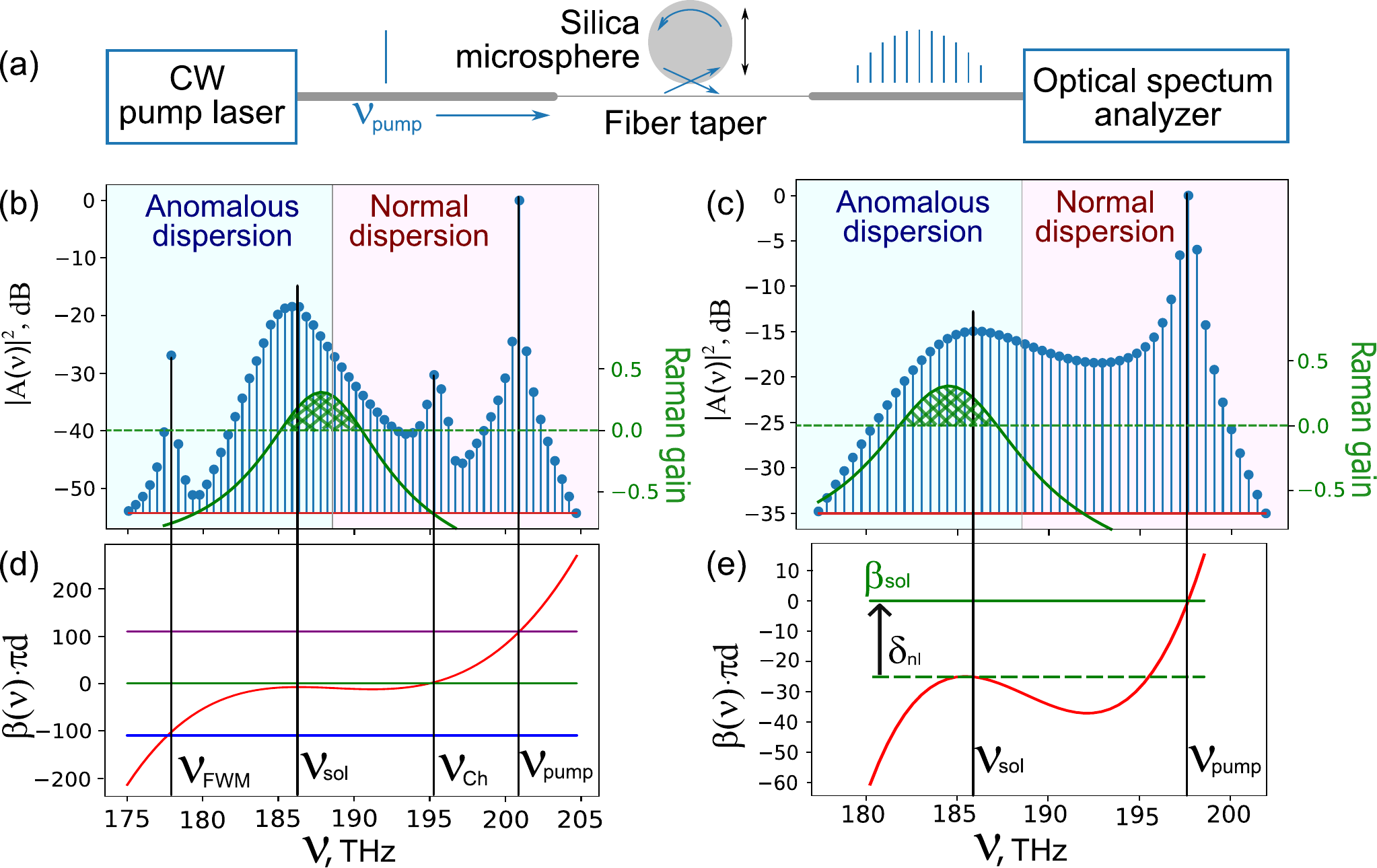}
\caption{
(a) Simplified scheme of the considered system. The spectra of non-synchronized (b) and synchronized (c) solitons for the pump at $\nu_{pump} = 200.93$ THz and $\nu_{pump} = 197.63$ THz, respectively. Raman gain profile (green lines in (b, c)); shading shows the region of positive gain. (d, e) Graphical solution of new resonant frequency generation.
The red curves are the linear waves dispersion in the reference frame moving with the soliton. The crossing of these curves with  the green, blue and purple lines corresponds to Cherenkov and FWM resonance conditions (Eq. (\ref{FWM condition})). Black vertical lines are the guides for eye showing that the resonant conditions predicts well the positions of the spectral lines associated with Cherenkov radiation and FWM. The dashed green lines in (e) is for the resonant condition when the nonlinear shift of the soliton propagation constant is neglected. }
\label{fig1}
\end{figure}

We have performed numerical simulation for a silica microsphere with diameter $d=140~{\mu}m$  optimized to achieve required dispersion (scheme is shown in Fig.~ \ref{fig1}(a)).  We demonstrate that by changing the CW pump frequency, it is possible to excite non-synchronized Raman solitons as well as the synchronized ones. The synchronized solitons are supported by both the Raman gain and the Cherenkov synchronism enabled energy transfer from the pump to the soliton. We also show that the synchronized Raman solitons can be controlled by a weak CW signal. 

To describe the dynamics of the light in the microspheres we adopt the widely used Lugiato-Lefever equation (LLE) \cite{lugiato1987spatial} written for a slowly varying amplitude of the mode resonantly coupled to the optical pump (for example, through a fiber taper). The LLE can be generalised to account for the higher-order dispersion and the Raman effect \cite{wang2018stimulated, anashkina2022thermo, JSTQE2024}:
\begin{eqnarray}
    \left( i t_R\frac{\partial }{\partial t}  + \pi d\hat \beta + \gamma\pi d ((1-f_R)|E|^2  + i\alpha  - \delta_0 \right) E -  i\sqrt{\theta}E_{in}+ \nonumber \\
     + \gamma\pi d\frac{f_R \left( T_1^{2} + T_2^{2} \right) }{T_1 T_2^2} E \int_0^\infty  e^{-\frac{s}{T_2}}\sin(\frac{s}{T_1}) |E(\tau-s, t)|^2 ds  = 0  \label{main_eq}
\end{eqnarray}
where $t$ is the slow time describing evolution of the
field envelope over multiple round trips, $\tau$ is the fast time describing the field envelope over a single round trip, $t_R = 1/FSR$ is the round-trip time, $ \hat \beta= \sum_{n=2}^{N} \frac{\beta_n}{n!} \big(i\frac{\partial}{\partial \tau})^n $ is the dispersion operator,
$\gamma$ is the Kerr nonlinear coefficient, $f_R$, $T_1$ and $T_2$ are the strength and the characteristic times of the Raman response, $\alpha$ is the effective losses and $\delta_0$ is the detuning of the pump from the resonance.
In our case it is sufficient to keep only three terms in the operator $\hat \beta$ to describe the dispersion of the waves with good accuracy. 
Thus we set $N=4$ and take the coefficients $\beta_2 = 10.49~ps^2/km$, $\beta_3 = 0.081~ps^3/km$, and $\beta_4 = -3.8\cdot 10^{-4}~ps^4/km$ calculated for the fundamental modes at the reference frequency $\nu_0 \approx 205.34~THz$. 
The field losses can be estimated as  $\alpha = \frac{(2\pi)^2R}{\lambda_0 Q}$ where the $Q$ - factor is $Q = 7\cdot 10^7$, $t_R = 2.12\, ps$, and $\theta$ is the coupling coefficient ($\theta \approx 2.6\times 10^{-5}$ at critical coupling). 
The coefficients characterising the Raman effect are  $f_R = 0.18,\: T_1 = 12.2\, fs, \: T_2 = 32\, fs$ \cite{wang2018stimulated}. Here, we consider different pump frequencies close to $\nu_{0}$.

It is convenient to measure the detuning of the pump from the resonance in the units of the losses and the power of the mode in characteristic nonlinear frequency shift. Therefore we introduce dimensionless detuning $\Delta_0 = \frac{\delta_0}{\alpha}$, dimensionless pump $f = E_{in} \sqrt{\frac{\pi \theta d \gamma (1-f_R)}{\alpha^3}}$, and dimensionless field intensity $|A|^2 = |E|^2 (1-f_R)\frac{\gamma\pi d}{\alpha}$. In all simulations presented in the paper  $\Delta_0 =  4$, $f = 3$, which corresponds to dimensional frequency detuning of $7.8$ MHz and pump power of $1.7$ mW at critical coupling. For simulations of (\ref{main_eq}) we have used a home-made code based on well-known Split-Step Fourier method.

We consider the case where the CW pump lies in the frequency of the normal dispersion and thus produces a dynamically stable plane wave propagating in the microsphere. However, for the properly chosen parameters, this wave produces Raman gain (amplification of the linear excitations after subtraction of the linear losses) for lower frequencies belonging to the anomalous dispersion range. For our parameters, the frequency dependence of the Raman gain is shown in Fig.~\ref{fig1}(b, c). One can see that there is a frequency range (shown in the figure by shading ) where the effective amplification is positive. The modes within this frequency range grow in time, and normally the fastest growing mode suppresses the others.  However, in our case the growing mode is in the range of anomalous dispersion and thus experiences modulation instability. At the nonlinear stage, this process results in the formation of a train of Raman solitons \cite{anashkina2022thermo} or Raman Turing patterns \cite{JSTQE2024}). 

The spectra of the stationary solitons are shown in Fig.~\ref{fig1} for the pump frequencies  $\nu_{pump} = 200.93$ THz (b) and $\nu_{pump} = 197.63$ THz (c). It is seen that the spectra are very different. Let us discuss the spectra and identify the observed spectral maxima.  

We start with the case $\nu_{pump}=200.93$ THz illustrated in Fig.~\ref{fig1}(b).  The numerical simulations show that in this case only one soliton forms in the resonator and the wide spectral maximum at $\nu_{sol} \approx 186$ THz can be identified as the soliton spectrum. As expected, the position of this spectrum maximum approximately coincides with the maximum of the Raman gain. 

To explain the whole spectrum, we need to take into account the resonant radiation emitted and scattered by the solitons. Indeed, apart from the spectral lines corresponding to the coherent pump, the field spectrum contains additional narrow spectral lines marked as $\nu_{ch}$ and $\nu_{FWM}$ in Fig.~\ref{fig1}(b). Let us discuss the physical origin of these lines. 

It is known that Cherenkov radiation occurs if the soliton moves at the velocity equal to the phase velocity of a linear wave. In other words, in the Fourier representation of the soliton there is a harmonic having the velocity equal to the phase velocity of a linear wave. To explain the other line at frequency $\nu_{FWM}$, we need to take into account that the pump not only causes the effective Raman amplification supporting the soliton but also can interact with the soliton through Four Wave Mixing (FWM) on Kerr nonlinearity. This effect is well known for the solitons propagating in nonlinear waveguides  \cite{PhysRevE.72.016619, cherenkov2017dissipative} and can easily be generalized for the system considered in this paper. 

The emission and the interaction of the dispersive waves of low intensity with the solitons can be described by a linear equation for the excitations nestling on the soliton background. The driving term in this equation contains the contribution from the effect of the higher-order dispersion on the soliton and the FWM of the soliton and the dispersive wave. The generation of new frequencies occurs if the driving term contains harmonics that are in the resonance with the eigenmodes of the system. The resonant condition for the Cherenkov radiation and for FWM is derived in  \cite{PhysRevE.72.016619, cherenkov2017dissipative} and in the reference frame moving with the soliton it reads 
\begin{equation}
    (\beta_{sol}+ \frac{\delta_{nl}}{\pi d}) + J[\beta(\nu_{pump})-(\beta_{sol} + \frac{ \delta_{nl}}{\pi d})] = \beta(\nu_{res}),
    \label{FWM condition}
\end{equation}
where $\beta(\nu)$ is the dispersion in the chosen reference frame,  $\beta_{sol}$ is the soliton propagation constant,  $\delta_{nl}$ is the soliton's nonlinear shift, $J = \{-1, 0, 1\}$ corresponds to different resonances with resonant frequencies $\nu_{res} = \{\nu_{FWM},\,\nu_{ch}\}$. $J = 0$ corresponds to the Cherenkov radiation, $J = \pm 1$ corresponds to the phase-sensitive and phase-insensitive  FMW processes.

Graphical solutions of the resonance condition (\ref{FWM condition}) is shown in Fig.~\ref{fig1}(d). The Cherenkov resonance occurs at the crossing of the dispersion of linear excitations (red line) with the dependency of the frequency of the soliton harmonics on their wave vectors. It is seen that resonant condition predicts the position of the spectral line marked as $\nu_{ch}$ in Fig.~\ref{fig1}(b) very precisely. The position of the resonant FWM of the soliton with the pump is also predicted very accurately, compare the position of the line $\nu_{FWM}$ with solution of the resonance condition (the crossing of the blue line with the red one) in Fig.~\ref{fig1}(d).   

The recoil from both, the Cherenkov radiation and the scattering of the coherent pump, affects the parameters of the soliton \cite{PhysRevE.72.016619} and this effect,  alongside with other effects like Raman self-frequency shift,  explains the deviation to the lower frequencies of the soliton frequency from the maximum of the Raman gain,  see  Fig.~\ref{fig1}(d). 

\begin{figure}[!t]
\centering
\includegraphics[width=0.85\linewidth]{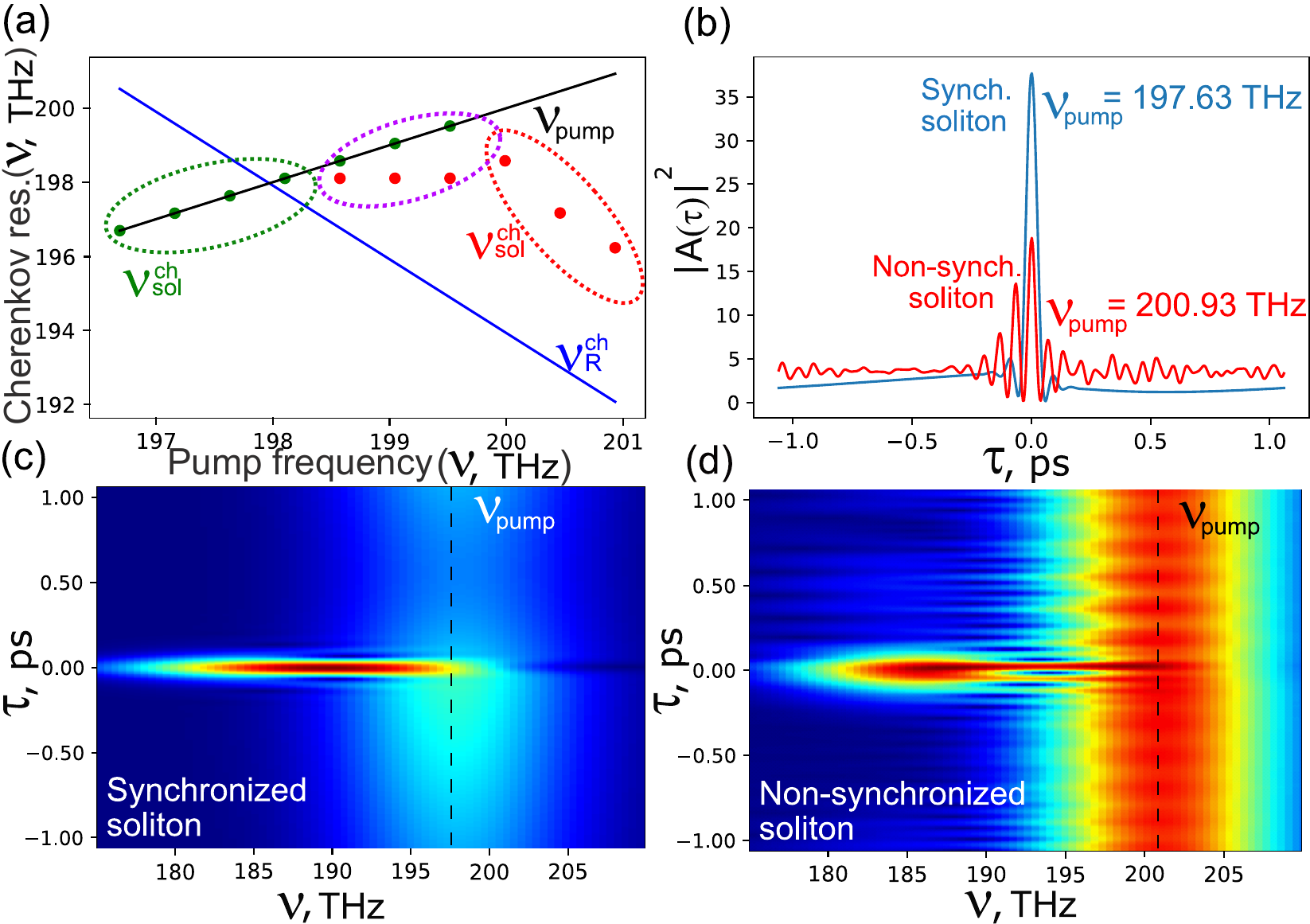}
\caption{The dots in panel (a) show the numerically found Cherenkov resonances of Raman solitons as function of the frequency of the pump supporting the solitons. The solid black line shows the dependency $\nu_{res}=\nu_{pump}$. The blue line is the position of the Cherenkov resonance calculated by the resonant condition under the assumption that the soliton frequency coincide with the maximum of the Raman gain. The green and the red ovals encircle the dots corresponding the synchronized and non-synchronized solitons. The dots in the magenta oval mark the Cherenkov frequencies for the soliton quasi-periodically  switching from the synchronized (green dots) to non-synchronized (dots) states. (b) The temporal distributions of the field intensities for synchronized (blue line)  and non-synchronized (red line)  solitons. XFROG traces of these two cases are shown in panels (c), (d). 
\label{fig2}}
\end{figure}

The resonance condition (\ref{FWM condition}) links the Cerenkov radiation frequency to the soliton frequency. One can extract the soliton frequency directly from numerical simulations and then calculate the Cherenkov frequencies. The Cherenkov frequencies found by this technique  are shown by dots in Fig.~\ref{fig2}(a). The Cherenkov frequency can also be calculated by approximating the soliton frequency from the Raman gain maximum. This calculation method gives the dependency of the Cherenkov resonance on the pump frequency shown by the blue curve in Fig.~\ref{fig2}(a). One can see that this method qualitatively fits the data extracted from direct numerical simulations for the pump at $\nu_{pump}=199.99$ THz, $\nu_{pump}=200.46$ THz and  $\nu_{pump}=200.93$ THz (dots in the red dashed oval). For these frequencies, the observed discrepancy is well explained by the deviation of the soliton frequency from the Raman gain maximum.

However, this is not so for the other pump frequencies. For the lower pump frequencies, the Cherenkov resonance is approaching the pump frequency and eventually the Cherenkov resonance merges with the  pump frequency (the dots on the black line in Fig.~\ref{fig2}(a)). The coincidence of the Cherenkov resonance and the pump frequency means that the soliton is phase-locked to the pump. These solitons will be referred to as synchronised solitons, while the other solitons will be referred to as non-synchronised. Let us remark that, as it follows from translational symmetry of Eq.(\ref{main_eq}),  the location of the synchronized soliton (on "fast time" axis) is not fixed. The translational symmetry can be broken, for example, by pulsed pump as it is done in \cite{li2023ultrashort}.  

The typical spectrum of the field of the synchronized Raman soliton is shown in Fig.~\ref{fig1}(c). It is seen that the narrow spectral line corresponding to the Cherenkov radiation and the scattered waves disappeared. Graphical solution of the resonance condition is shown in  Fig.~\ref{fig1}(e). The predicted position of Cherenkov resonance fits to the frequency of the pump perfectly.

Let us mention that the synchronized soliton is quite intense and so to get a good agreement we need to take into account the nonlinear shift of the soliton propagation constant from the value predicted by the dispersion characteristic for the linear waves. So, to find the Cherenkov synchronism we extracted the soliton propagation constant from the numerical simulations. The resonant conditions solution for the soliton propagation constant approximated by the propagation constant of the linear wave is shown by the dashed line and it is seen that the discrepancy is large. It is also worth mentioning that, in an agreement with the resonance conditions, there is no scattering of the pump on the soliton. This explains the spectrum of the synchronized solitons shown in Fig.~\ref{fig1}(c).

To demonstrate the phase locking of Raman solitons to the pump it is instructive to compare cross-correlation frequency resolved optical gating (XFROG) traces Fig. \ref{fig2}(c)(d) and the temporal filed distributions Fig.\ref{fig2}(b) for the synchronized and non-synchronized solitons.  Synchronized solitons are always in the same phase in respect to the pump and therefore no interference fringes are seen on the XFROG trace of in the temporal distribution of the field. It is also seen on the XFROG that the solitons cause significant depletion of the pump. In the same time, if the solitons are out of the Cherenkov synchronism then the total field (the pump and the soliton) is evolving in time with deep interference fringes varying with the slow time. These fringes are also seen well in the XFROG trace (Fig.~\ref{fig2}(d)). 

The intensity distributions (on "fast time") for the synchronized and non-synchronized solitons are shown in Fig.~\ref{fig2}(b) where it is seen that the soliton synchronized to the pump has higher intensity and shorter duration than the non-synchronized soliton. Correspondingly, the spectrum of the synchronized solitons is wider containing larger number of phase-locked harmonics, compare the spectra shown in Fig.~\ref{fig1}(b, c).

The fact that the relative phase of the synchronized soliton and the pump is not changing in slow time means that coherent parametric processes contribute to the energy exchange between the soliton and the pump. This increases the intensity of the synchronized solitons significantly. Since the solitons are pumped mostly through inverse Cherenkov effect, the frequency of the solitons does not follow the maximum of the Raman gain but is defined by the Cherenkov synchronism, see Fig.~\ref{fig2}(a).

It is interesting to mention that apart from synchronized and non-synchronized solitons there may be solitons that can be seen as quasi-periodic switching from the synchronized to non-synchronized states. The switching process is much faster then the life time in each of the quasi-stationary states. The Cherenkov frequencies of the quasi-stationary states are shown in Fig.~\ref{fig2}(a) in dashed magenta oval by green (synchronized meta-stable state) and red (non-synchronized meta-stable state) dots.

\begin{figure}[!t]
\centering
\includegraphics[width=0.85\linewidth]{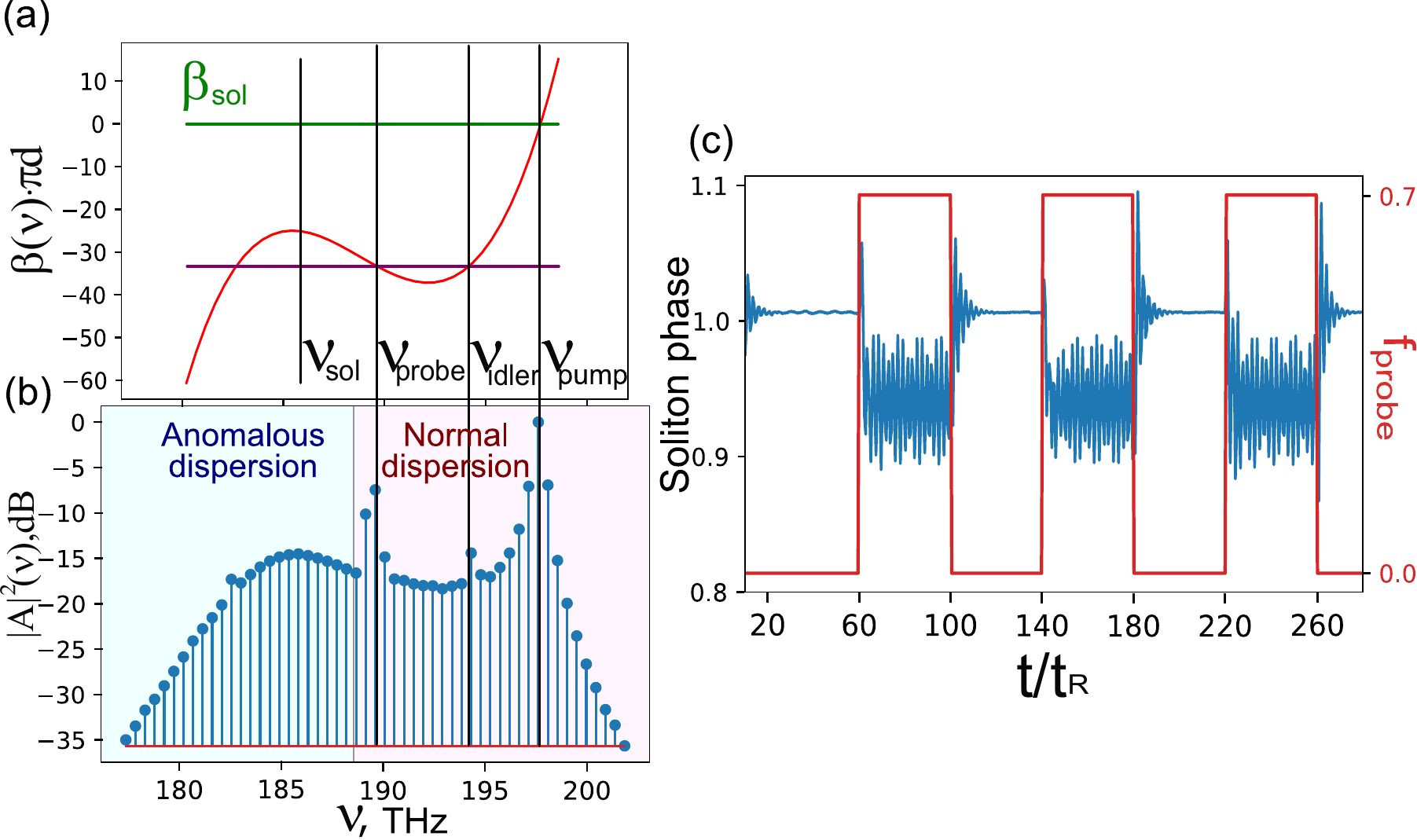}
\caption{
(a) The graphical solution of the resonant condition in the presence of the probe. The meaning of the lines is the same as in \textcolor{blue}{Fig.~}\ref{fig1}(d). (b) The spectrum of the synchronized soliton in the presence of the probe.  (c) The mutual phase of the field in the resonator and the pump calculated at the position of the soliton intensity maximum. The red line shows the intensity of the probe signal varying from zero to $f_{probe} = 0.7\approx 0.23 f_{pump}$, the frequency of the probe is $\nu_{probe} \approx 189.61$ THz}.
\label{fig3}
\end{figure}

Finally, we also checked that the phase of the soliton with respect to pump can be controlled by additional weak coherent probe. As it is discussed above, a FWM of the soliton with dispersive waves can generate new frequencies and the recoil from this mixing changes the parameters of the soliton. We take a synchronized soliton and affect it by a weak CW probe at the frequency $\nu_{probe}$. The resonant condition predicts the appearance of the new frequencies that are indeed present in the spectrum of the soliton, see Fig.~\ref{fig3}.

As it is mentioned above, the relative phase between the soliton and the pump is fixed. We use a probe beam switched on and off to show that this relative phase can be changed by the probe. To do this, we watch the temporal evolution of the field phase at the soliton center with respect to the phase of the holding beam (the pump). This evolution is shown in Fig.~\ref{fig3}(c). It is clearly seen that when the probe is on it shifts the soliton phase. The oscillations of the phase appear because of the interference of the probe and the soliton. When the probe is switched off, the phase always returns to the same value. This proofs the stable phase locking of the soliton to the pump and the possibility to control the relative phase of the soliton by the probe.

To conclude, we demonstrated that the Raman solitons  forms in silica microspheres can be in and out of the Cherenkov synchronism with coherent pump. In the synchronous regime the energy is transferred from the pump to the solitons via two channels: through Raman gain and through Cherenkov synchronism. This increases the intensity of the soliton and makes the spectrum of the generated frequency comb wider. It is also demonstrated that the phase of the soliton is locked to the phase of the pump and that this phase can be efficiently controlled by a weak CW probe. We believe that the reported findings can be of interest from the point of view of soliton dynamics and can be used to improve the parameters of the frequency combs generated in microresonators.

\begin{backmatter}
\bmsection{Funding} 
 This work was supported by the Ministry of Science and Higher Education of Russian Federation, goszadanie no. 2019-1246. and by Priority 2030 Federal Academic Leadership Program,  Ministry of Science and Higher Education of the Russian Federation (075-15-2022-316);  Russian Science Foundation (20-72-10188-P) 
 
\bmsection{Acknowledgements} 
ANO and AVY acknowledge the financial support from the Ministry of Science and Higher Education of Russian Federation, goszadanie no. 2019-1246 and from Priority 2030 Federal Academic Leadership Program. The work of E.A.A. was supported by the Center of Excellence ``Center of Photonics” funded by the Ministry of Science and Higher Education of the Russian
Federation, 075-15-2022-316 (calculation of dispersion coefficients) and by the RSF 20-72-10188-P (partial development and analysis of the model).

\bmsection{Disclosures} The authors declare no conflicts of interest.

\bmsection{Data availability} Data underlying the results presented in this paper are not publicly available at this time but may be obtained from the authors upon reasonable request.

\end{backmatter}

\bibliography{sample}

\bibliographyfullrefs{sample}

\end{document}